\documentclass[oneside,onecolumn,superscriptaddress,showpacs]{revtex4}
\usepackage{graphics}
\begin{document}

\title{Testing of various parametrizations of effective
nucleon-nucleon interaction in elastic and inelastic scattering of
polarized protons from $^{12}$C }

\author{M.S.~Onegin}
\affiliation{188300 Petersburg Nuclear Physics Institute, Gatchina, Russia}
\author{A.V.~Plavko}
\affiliation{194223 State Polytechnic University of St. Petersburg, Russia}

\date{\today}

\begin{abstract}
The parametrization schemes of the effective nucleon-nucleon
interaction proposed by Geramb and by Nakayama and Love are compared in
their ability to reproduce the inelastic scattering observables of
polarized protons from $^{12}$C with the excitation of the $1^+$ level
with $T=1$.  The off-shell behavior of the effective $t$ matrix is
shown to be important.  Density dependence of the effective interaction
is also analyzed.  \end{abstract}


\maketitle


\section{Introduction}

The inelastic scattering of protons from nuclei is a useful tool of
investigating nuclear properties. It is necessary to know the effective
$NN$ interaction between the projectile proton and nucleons in the
nucleus to extract information about the properties of its excited
states. The problem of the modification of the free $NN$ interaction in
the nuclear medium has been studied in many papers (see
e.g.~\cite{Brieva,Geramb,Nakayama,Kelly,Sammaruca}). The effective $NN$
interaction can be used in local  density approximation for the description of
elastic~\cite{Karatag,Geramb1} or inelastic~\cite{Geramb1,Kelly1,Stephenson} scattering.

For proton energies in the range from 100 to 400 MeV, the main
modification of the free $NN$ scattering reaction $t$ matrix in the
nuclear medium is due to the Pauli exclusion principle for scattering
in occupied states and due to the influence of the mean nuclear field
inside the nuclear matter. It is the g-matrix function that is used to
quantitatively describe this influence.  The g matrix is a solution of
the Bethe-Goldstone equation~\cite{Mahaux} - the equivalent of the
Lippman-Swinger equation for the t matrix in the nuclear medium. There
are two approaches to the solution of this equation which are
different: one is based on coordinate space~\cite{Mahaux,Geramb} and
the other one on impulse space~\cite{Nakayama,Geramb2,Amos}.  The
parametrization by von Geramb~\cite{Geramb} was especially successful.
It provides the so- called Paris-Hamburg (PH) effective interaction
which describes elastic and inelastic scattering uniformly well for
nuclei of different atomic weight.

Later a new parametrization of the effective interaction was proposed by Nakayama and
Love (NL)~\cite{Nakayama}. In this work the authors achieved some
success in the description of high-spin stretched states in $^{28}$Si.
In work~\cite{Kelly} it was claimed that the NL construction better
reproduces the low density behavior of the tensor component of the
effective interaction as compared with the free $NN$ interaction and it
is important for the success in the case of stretched states. In the
present paper we hold the converse opinion - that the similarity
between the NL interaction and the free $NN$ interaction is a disadvantage of the NL
construction.  We compare the predictions of inelastic scattering observables based on these two
parametrizations of the effective interaction with the experiment and show that the Geramb
interaction reproduces them better.  The reason for the success of the Geramb forces is also
analyzed.

\section{Elastic scattering}

The optical potential of the nucleon in nuclear matter can be
calculated in the first order microscopically~\cite{Karatag}. The
Geramb and NL g matrices lead to optical potentials considerably
different in their radial form. They are sensitive only to the
isoscalar components of the g matrix. The elastic observables,
differential cross sections   ($d\sigma /d\Omega $), analyzing power
($A_y$), and the spin-rotation parameter ($Q$) calculated using these
potentials are represented in Fig.1.

\begin{figure}
\scalebox{0.5}{\includegraphics{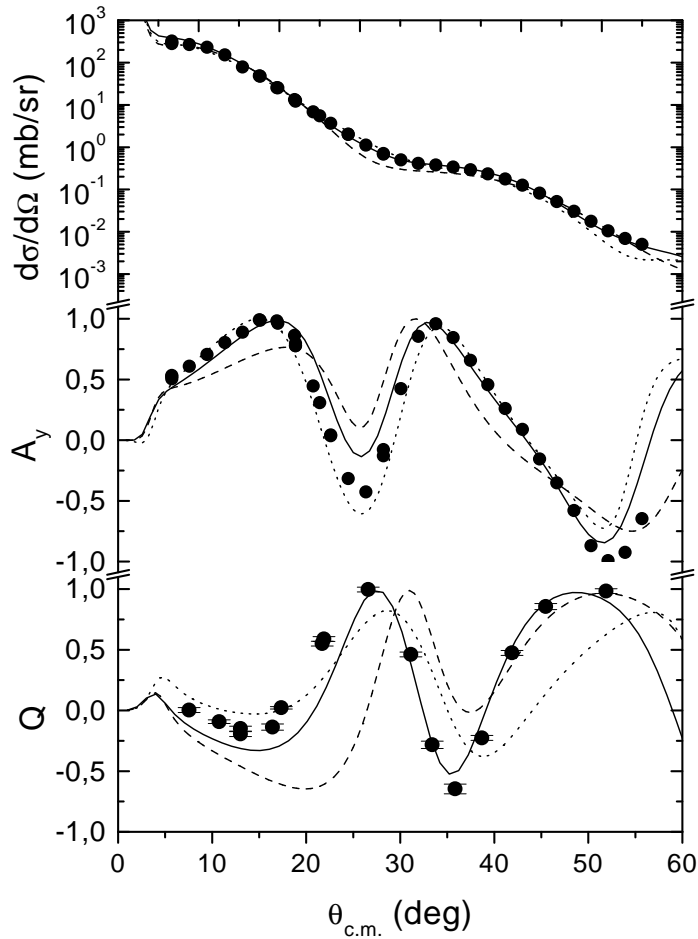}}
\caption{\label{fig1}Comparison of the calculated elastic scattering
observables $d\sigma /d\Omega , A_y $ and $Q$ with the experiment. The
experimental data are taken from ~\cite{Comfort,Meyer}. Calculations
(DWBA-91): the solid lines show the microscopical calculations using the
Geramb interaction, dashed lines show the same, using the NL interaction,
and the dotted lines indicate calculations, using the phenomenological WS
potential.}
\end{figure}

As is seen from the comparison
with the experimental data~\cite{Comfort}, the Geramb g matrix better
reproduces the experiment. The reason for the preference of the Geramb
forces over the NL ones in the isoscalar channel was explained in
work~\cite{Kelly}. As different optical potentials influence the
calculated inelastic observables,   further we use the phenomenological
optical potential~\cite{Meyer} for the calculation of distorted waves
for the inelastic channel.  This potential satisfactorily reproduces
the experimental observables (see Fig.1) of elastic scattering.

\section{Construction of the effective interaction}

Von Geramb, Nakayama and Love used a Bethe-Goldstone type equation for
the infinite nuclear matter to obtain the $g$-matrix interaction of a
free nucleon with a nucleon in the nuclear medium. The nuclear matter
is described by non-interacting Fermi gas with the Fermi impulse $k_F$.
The difference was that Nakayama and Love solved this equation in an
impulse representation while von Geramb worked in coordinate space.

Lets assume ${\bf k}_0$ and ${\bf k}$ to be relatively initial and
final relative momenta of two nucleons and ${\bf K}$ as their
center-of-mass momentum.  The Bethe-Goldstone equation then will have
the following form~\cite{Haftel}:

\begin{eqnarray}
 G({\bf K}|{\bf k},{\bf k}_0)=V({\bf k},{\bf k}_0)+
\int d^3k \frac{V({\bf k},{\bf k}')Q({\bf K},{\bf k'})
G({\bf K}|{\bf k'},{\bf k}_0)}
{E({\bf K},{\bf k}_0)-E({\bf K},{\bf k}')+i\epsilon},
\label{BG}
\end{eqnarray}
where $Q({\bf K},{\bf k'})$ satisfies the Pauli principle condition:
$Q({\bf K},{\bf k}')=1$ for $|{\bf K}+{\bf k}'| > k_F$ and $Q({\bf
K},{\bf k}')=0$ for $|{\bf K}\pm {\bf k}'| < k_F$. The function $E({\bf
K},{\bf k})$ in the denominator denotes the sum of single-particle
energies of the two nucleons. Usually the operator $Q$ in the equation
~(\ref{BG}) is replaced by its angle averaged value~\cite{Haftel}
$$ \overline{Q}(K,k') =\frac{1}{4\pi }\int d\hat k' Q({\bf K},{\bf
k}').$$
The energy denominator in the equation~(\ref{BG}) is also substituted
by its angle averaged form though the prescription of this averaging is
different in the Geramb~\cite{Geramb} and NL~\cite{Nakayama} papers.
This difference was discussed in~\cite{Kelly} and we do not analyze it
here because it is absent for the case of $k_F=0$. As we show later,
the $g$ matrices obtained by von Geramb and Nakayama and Love are also
different in this particular case and it is the main reason for the
differences in predicted inelastic observables we consider based on
these two interactions.

With the angle averaged Pauli operator $Q$ and
the denominator the solution of the Bethe-Goldstone equation does not
depend any more on the direction of the vector ${\bf K}$.  The
remaining angular dependent part of the function $G$ can be expanded
using partial-wave decomposition

\begin{eqnarray}
G(K|{\bf k},{\bf k}_0)=\frac{2}{\pi}\sum_{{JST\atop
LL',M}}i^{L-L'} G^{JST}_{L'L}(K|k,k_0) T_{LSJ,M}(\hat k_0)^*
T_{L'SJ,M}(\hat k) P_T,
\label{expan}
\end{eqnarray}
where the spin-angle function $T_{LSJ}$ is an eigenstate of two
nucleons with the total angular momentum $J$, total spin $S$ and the
total orbital angular  momentum $L$. The operator $P_T$ is a projector
on the state of two nucleons with the total isospin $T$. The
substitution of the decomposition~(\ref{expan}) by ~(\ref{BG}) leads to
a coupled-channel integral equation for the matrix elements
$G^{JST}_{L'L}(K|k,k_0)$. It can be numerically solved for each channel
with the quantum numbers $JST$. This method was developed by Nakayama
and Love.

Von Geramb solved a related Bethe-Golstone equation for the correlated
wave function (WF) of two nucleons $\psi({\bf r},K,{\bf k}_0)$. This
equation has the following form~\cite{Geramb1}:

\begin{eqnarray}
\psi = \phi + G_Q^{(+)} V \psi,
\label{RBG}
\end{eqnarray}
where $V$ is the free $NN$ interaction, $\phi$ is the plane wave
function  of the relative $NN$ motion and $G_Q^{(+)}$ is the Green
function of two nucleons in the nuclear matter with outgoing wave
boundary conditions

\begin{eqnarray}
G_Q^{(+)} =\sum_{|{\bf k}_1|,|{\bf k}_2|>k_F}
\frac {Q({\bf k}_1, {\bf k}_2)}{E-e_1-e_2+i\epsilon}.
\nonumber
\end{eqnarray}
Here $Q$  is the Pauli projecting function for two nucleons in
intermediate states with the wave vectors ${\bf k}_1$ and ${\bf k}_2$,
$e_1, e_2$ are single particle energies of nucleons in the nuclear
medium.

 The $g$ matrix and $\psi$ are connected with the following
expression:

\begin{eqnarray}
<\phi |G|\phi >=<\phi|V|\psi >.
\nonumber
\end{eqnarray}
This equation allows to interpret the $g$ matrix as an effective $NN$
interaction  in the nuclear matter. The numerical solution of
equation~(\ref{RBG}) is performed with the expansion of  the WF $\psi$
over the multipoles by

\begin{eqnarray}
\psi ({\bf r},K,{\bf k}_0)=\sum_{JLL'\atop ST} (4\pi(2L+1))^{\frac
{1}{2}}i^L u^{JST}_{L'L}(r,k_0,K) \\
\times\, T_{L'SJ,M}(\hat r,\hat k_0) <L0SM|JM>
P_S P_T \nonumber
\end{eqnarray}
and with the use of well-known techniques for the solution of wave
equations.

In both considered cases the desired local complex energy-
and density-dependent effective interaction has the following operator
form:

\begin{eqnarray}
V(r)=\sum_{ST} V^C_{ST}(r)P^SP^T+\sum_T V^{LS}_{T}(r) {\bf L}\cdot {\bf
S}+ \sum_T V^T_{T}(r) S_{12}.
\label{force}
\end{eqnarray}
Here $L$ is a total orbital angular momentum operator and $S_{12}$ is
a usual tensor spin operator. The antisimmetrized matrix element of
this force between two nucleon states with the relative wave vectors
${\bf k}_0$ and ${\bf k}$ in the initial and final channels
respectively has the following form~\cite{LF,Kelly}:

\begin{eqnarray}
\tilde t({\bf k},{\bf k}_0) = \sum_{ST}\tilde t^{\,C}_{ST} P_SP_T+
i\sum_T\tilde t^{\,LS}_T ({\bf \sigma _1}+{\bf \sigma _2})\cdot \hat
{\bf n} P_T \label{qforce} \\
-  \sum_T [\tilde V^{\,T}_T(q)S_{12}(\hat
{\bf q})-(-)^{1+T}\tilde V^{\,T}_T(Q)S_{12}(\hat {\bf Q})]P_T,
\nonumber
\end{eqnarray}
where ${\bf q}={\bf k}_0-{\bf k}$, ${\bf Q}={\bf k}_0+{\bf k}$,
$\hat {\bf n}=\hat {\bf q} \otimes \hat {\bf Q}$, and where
\begin{eqnarray}
\tilde t^{\,C}_{ST}(q)=\tilde V^{\,C}_{ST}(q)-(-)^{S+T}
\tilde V^{\,C}_{ST}(Q), \label{fourie} \\
\tilde t^{\,LS}_{T}(q)=\frac{1}{4}qQ[\tilde V^{\,LS}_{T}(q)+
(-)^{1+T}\tilde V^{\,LS}_{T}(Q)].
\nonumber
\end{eqnarray}
The functions $V^{\,C,LS,T}(q)$ are the Fourier transforms of the
corresponding radial functions in~(\ref{force})~\cite{LF,Kelly}.

When Nakayama and Love were constructing various components of radial
dependent functions  in~(\ref{force}) from the matrix elements
$G^{JST}_{L'L}(K|k,k_0)$, they used these matrix elements only on-shell
($k=k_0$). Using some simplifications, the wave vectors $K$ and $k$ can
be evaluated via the wave vector of the projectile $k_1$ and the Fermi
momentum $k_F$. If $\theta$ is the scattering angle in the c.m. frame,
then the momentum transfer $q$ is equal to
$$ q=2k_0sin\frac{\theta}{2}.$$
The spin-dependent antisimmetrized part of the effective interaction
in the central channel for example can be obtained through the
following equation:

$$ G_1^C(q,k_1,k_F)=\frac{1}{2\pi ^2}\sum_{LJ}(2J+1) G^{J1T}_{LL}(K|k,k)
P_L(cos\theta)\frac{1}{3}P_{S=1}.$$
The summation in this equation over $L$ is restricted by the Pauli
principle. This interaction can be further expanded into isoscalar and
isovector parts. Nakayama and Love identified then the isovector
component with the following combination of $NN$ $t$-matrix elements of
equation~(\ref{qforce}):

\begin{eqnarray}
  G_1^C(q,k_1,k_F)=\eta (\tilde t^{\,C}_{00}(q,k_1)-\tilde
t^{\,C}_{01}(q,k_1) -\tilde t^{\,C}_{10}(q,k_1)+\tilde
t^{\,C}_{11}(q,k_1))/16,
\end{eqnarray}
where $\eta$ is a Jacobian of transformation from the $NN$ center of
mass to the $NA$ center of mass~\cite{LF}. In the same manner Nakayama
and Love expressed other components of effective
interaction~(\ref{qforce}) via a corresponding combination of
$g$ matrix elements.  The desired effective interaction was obtained in
a fitting procedure for different energies of the projectile and Fermi
momentum $k_F$.

Von Geramb, however, used a different fitting procedure.
Following~\cite{Brieva} he identified the $g$ matrix effective
interaction with the averaged two nucleons free $NN$ interaction by
using plane incoming and correlated outgoing wave functions. This
procedure reproduces in general radial correlations of two-body WF in
the nuclear matter carried by $\psi$:

\begin{eqnarray}
V_{ST}^C(r)=\frac{\sum_J \hat J (LJST|t|LJST)}{\hat S \hat L},
\label{vger}
\end{eqnarray}

where
\begin{eqnarray}
 (L'JST|t|LJST)&& \nonumber \\
 &=& \frac{\sum_{L"}\int_{|{\bf p}|\le
k_f}d^3pj_{L}(k_0r)V^{JST}_{L"L}(r) u^{JST}_{L'L"}(r,k_F,k,K)}
{\int_{|{\bf p}|\le k_f}d^3p j_{L'}(k_0r)j_{L}(k_0r)}.
\end{eqnarray}

The central component of the effective interaction obtained according
to~(\ref{vger}) depends on the orbital momentum $L$. Additional
averaging was performed with plane wave states. The effective
interaction obtained in this manner was approximated in terms of Yukawa
form factors.

\section{Comparison of effective isovector interactions}

To compare the Geramb and NL interactions, we reproduce in Fig.2 the
real and imaginary parts of the central isovector spin-dependent,
isovector spin-orbit and direct isovector tensor components of the $t$
matrix (Eq.~(\ref{qforce})) for $k_F=0$. The spin-orbital component of
the force $\tilde \tau ^{LS}$ is expressed through $\tilde t ^{LS}$
(Eq.~(\ref{fourie})) in the following equation:

$$ \tilde \tau ^{LS}(q)=-\frac{1}{qQ}\tilde t ^{LS}(q)$$

Fig. 2 also shows the $t$ matrix obtained from the Franey and Love
(FL) parametrization of the free NN interaction~\cite{FL}.

\begin{figure}
\scalebox{0.5}{\includegraphics{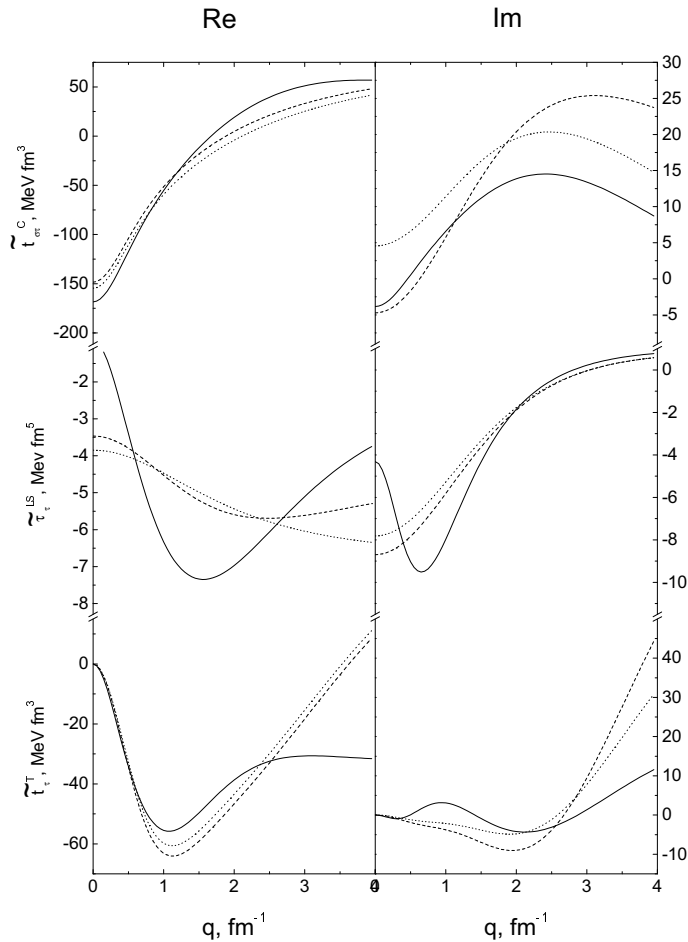}}
\caption{\label{fig2}The $q$ dependence of different components of the
effective $t$-matrix interaction (real - the left-hand side and imaginary -
the right-hand side respectively).  The solid lines show the Geramb
interaction, dashed lines indicate the NL, and the dotted lines show
the FL free interaction.}
\end{figure}

As is seen
from the comparison of the real parts of different components of the NL
and FL interactions, they are similar, while the Geramb interaction is
considerably different from both.  The imaginary part of the NL
interaction is also more similar to the FL one and both of them are
notably different from the Geramb interaction.  As in
work~\cite{Kelly}, we believe that this is due to an approximate
taking into account of the off- shell behavior in the Geramb scheme of
the approximation of the $g$ matrix by the local Yukawa form potential
and the lack of it in the NL scheme.

\section{Description of inelastic scattering of polarized protons with
the excitation of the $1^+, T=1$ level in $^{12}$C using various
effective interactions}

The excitation of the level $1^+, T=1$ with $E^*=15.11$~MeV in
$^{12}$C is produced with those isovector spin-dependent components of
the force ~(\ref{force}) that we have investigated in the above
section. We used the Cohen-Kurath wave function for the description of
the excited state~\cite{Cohen}. It satisfactorily reproduces the
$B(M1)$ value of this level as well as the momentum dependence of the
transverse formfactor of $(e,e')$ scattering~\cite{Karatag}. The
calculations of the inelastic scattering observables were performed in
the framework of Raynal's program DWBA-91~\cite{DWBA}. It takes into
account the antisimmetrization of the WF of the incident proton and
active  nucleons in the nuclear. We tested the Geramb and NL forces as
effective interactions. Their comparison with the experiment is
presented in Fig.3. The experimental differential cross-section
($d\sigma /d\Omega $) and part of the analyzing power ($A_y$) data were
obtained in work~\cite{Comfort}. The energy of scattered protons in
this experiment was 200 MeV. The remaining data of the analyzing power
, polarization ($P$) and the polarization transfer coefficient
$D_{NN'}$ were taken at 198 MeV from work~\cite{Opper}. $D_{NN'}$   is
only one normal-component polarization transfer observable.

\begin{figure}
\scalebox{0.5}{\includegraphics{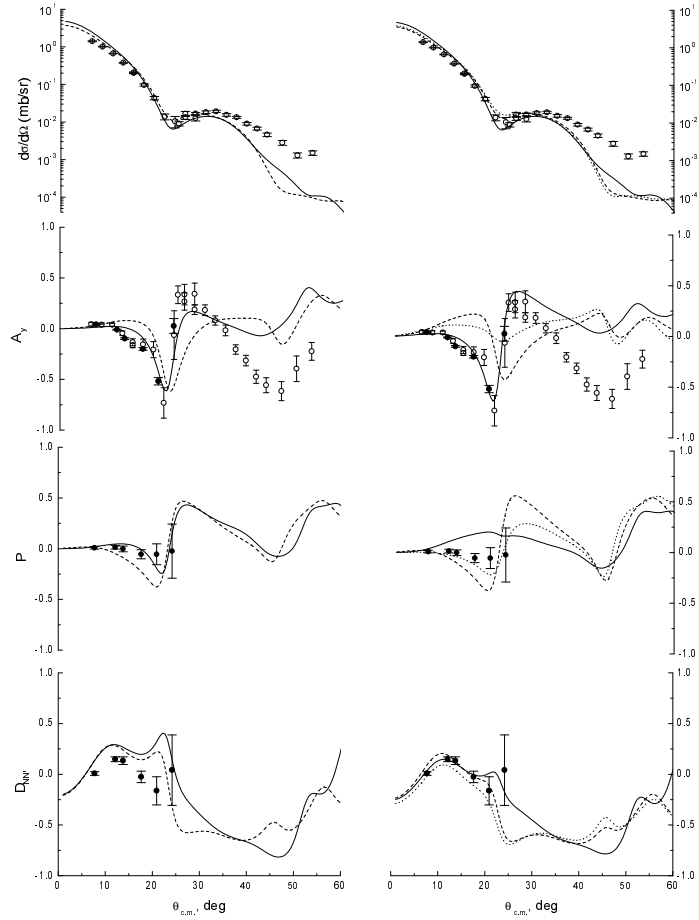}}
\caption{\label{fig3}Comparison of the calculated differential cross-section
($d\sigma /d\Omega$) , analyzing power ($A_y$), polarization ($P$) and
polarization transfer coefficient $D_{NN'}$  for the level $1^+$ in
$^{12}$C with the experiment ($E_p\simeq 200$ MeV). The open circles
indicate the data taken from~\cite{Comfort}, the solid curves show the
results from~\cite{Opper}. The left-hand side represents density
dependent calculations, the right-hand side shows those without density
dependence. The solid lines show the Geramb interaction, dashed lines
indicate the NL, and the dotted lines show the FL interaction.}
\end{figure}

Two cases were considered with the use of density dependence and
without it. They are presented on the left- and right-hand side of
Fig.3 respectively. For the case without density dependence, we have
also presented calculations with the use of the FL parametrization of
the free $t$ matrix of the $NN$ interaction. As can be seen from the
comparison of the density dependent calculations with the experiment,
the NL interaction describes the cross-section almost as well as the
Geramb interaction. The deterioration in the description of the
experiment with the increase of the scattering angle is also seen in
inelastic electron scattering, where the WF of Cohen and Kurath do not
describe properly the second maximum of the transverse form
factor~\cite{Karatag}. The behavior of the cross-section at small
angles is following the absolute square of the $t^C_{\sigma \tau}$
component which is very similar in both cases. The behavior of the
$A_y$ and $P$ is different for the Geramb and NL interactions. The
Geramb interaction better reproduces the experimental data. The shift
of the NL analyzing power towards large angles (see also
~\cite{Nakayama}) is the consequence of different behavior of $t^{LS}$
and $t^{T}$ versus $q$ in both parametrizations. This difference is
further amplified when the density independent interactions are used
(the right-hand side of Fig.3). Nearly the same situation is seen for
polarization. In both cases the use of the FL interaction in our
calculations leads to a similar behavior of the analyzing power and
polarization as with the NL interaction.

The  calculated polarization transfer coefficient is similar for both
the Geramb and NL interactions.  It overestimates the experimental data
in both cases.  Though this overestimation is absent for density
independent forces, this fact just reflects some uncertainty. On the
whole the DWBA calculations with the Geramb interaction better
reproduce the experimental data than with the NL and FL one.

\section{Conclusion}

We examine the Geramb and NL effective interactions in their ability
to reproduce the excitation of the isovector anomaly parity state $1^+$
in $^{12}$C with polarized protons. This excitation is driven with the
isovector spin-dependent components of the effective $NN$ interaction.
As is seen from the construction of the effective interaction by von
Geramb, it takes into account its off-shell behavior approximately. On
the other hand, the construction of the NL interaction completely
ignores the off-shell behavior of the $g$ matrix - the solution of the
Bethe-Golstone equation. We think that it is this advantage of the
Geramb (PH) interaction that allows to reproduce satisfactorily the
polarization observables of inelastic scattering. The elastic
sattering calculations using the PH model are especially successful in
the description of the elastic observables, including the
triple-scattering parameter $Q$ (Fig. 1). The good description of the
latter is rather remarkable for nonrelativistic theories.

{\begin{center}Acknowledgements\end{center}}

 The authors thank Prof. P. Schwandt for his tables of
effective NN interactions.



\end{document}